%% file: paper.tex
\documentclass[aps,pre,amsfonts,showpacs]{revtex4}
\usepackage{epsfig,amsopn}
\usepackage{graphicx}
\usepackage{psfrag}

\begin{document}
\def\be{\begin{equation}}
\def\ee{\end{equation}}
\def\bea{\begin{eqnarray}}
\def\eea{\end{eqnarray}}
\def\l{\label}

\title{Relaxation times of unstable states in systems with long range interactions}
\author{Kavita Jain$^1$, Freddy Bouchet$^2$ and David Mukamel$^1$}
\affiliation{$^1$ Department of Physics of Complex Systems, The Weizmann
  Institute of Science, Rehovot 76100, Israel \\
$^2$ Institut Non Lin\'eaire de Nice (INLN), CNRS, UNSA UMR 6618, 1361 route des lucioles, 06 560 Valbonne, France}
\date{\today}

\begin{abstract}
We consider several models with long-range interactions evolving
via Hamiltonian dynamics. The microcanonical dynamics of the basic
Hamiltonian Mean Field (HMF) model and perturbed HMF models with either global 
anisotropy or an  on-site potential are studied both analytically and
numerically. We find that in the magnetic phase, the initial zero 
magnetization state
remains stable above  a critical energy and is unstable below it.
In the dynamically stable state, these models exhibit
relaxation time scales that increase algebraically with the number
$N$ of particles, indicating the robustness of the quasistationary
state seen in previous studies. In the unstable state, the corresponding 
time scale increases logarithmically in $N$.
\end{abstract}
\maketitle


\section{Introduction}

In recent years, much work has been devoted to the understanding of
the statistical mechanics and the dynamics of systems with long
range interactions. In these systems the interaction potential
between two particles decays at large distances $r \gg 1$ as $V(r)
\sim r^{-\alpha}$ with $\alpha \leq d$, $d$ being the dimension of
the system. Examples of such systems include self-gravitating
systems, two-dimensional and geophysical vortices, non-neutral
plasma and systems describing wave particle interactions (Free
Electron Laser, CARL experiment etc.), and magnetic dipolar
systems (see \cite{Dauxois:2002} for reviews).

The long range nature of the interactions makes these systems
non-additive. Due to this property, at statistical equilibrium,
inequivalence between the microcanonical and the canonical
ensembles is a generic feature. This was first observed in
models of self gravitating stars
 \cite{Lynden-Bell:1968, Hertel:1971}, and then seen
in a number of other models ranging from the point vortex model
\cite{Smith:1990,Caglioti:1990}, plasma physics
\cite{Kiessling:2003}, self gravitating systems
\cite{Chavanis:2002b,Miller:1998}, two dimensional flows
\cite{Ellis:2002}, long range Hamiltonian models
\cite{Antoni:2002,Dauxois:2003} to simple spin models with mean
field interactions \cite{Barre:2001,Mukamel:2005}. A
classification of phase transitions and ensemble inequivalence in
generic long range systems \cite{Bouchet:2005c} has shown that
many possible types of behavior remain to be seen in specific
physical systems.

Beside these equilibrium peculiarities, the dynamics of systems
with long range interactions also present several new features.
For a large number $N$ of particles, these systems may exhibit
quasi-stationary states (QSS) \cite{Latora:1998,Yamaguchi:2004}
(in the plasma or astrophysical context, see for instance
\cite{Dubin:1999,Spitzer:1987}), very long relaxation time
\cite{Yamaguchi:2004}, vanishing Lyapounov exponents
\cite{Latora:1998,Firpo:1998a}, anomalous relaxation and diffusion
\cite{Latora:1999,Yamaguchi:2003,Pluchino:2004,Bouchet:2005a,Yamaguchi:2007}
and breaking of ergodicity \cite{Borgonovi:2004,Mukamel:2005}.
These features are a result of similar collective
(self-consistent) dynamics \cite{Del Castillo:2000} shared by
systems with long range interactions. In the limit of large
number of particles, such dynamics is well approximated by kinetic
theories
\cite{Spitzer:1987,Nicholson:1983,Dubin:2003,Chavanis:2000,Bouchet:2005a}
which to leading order in $1/\sqrt{N}$ describe Vlasov type
dynamics, and after much longer time, the relaxation towards
equilibrium is governed by Lennard-Balescu type dynamics.

In this paper, we consider the dynamics of systems with long range
interactions and analyze the time it takes for a system to relax
to its equilibrium state, starting from a thermodynamically
unstable state. The relaxation process is usually initiated by the
formation of droplets of the equilibrium state which is followed by
a coarsening process \cite{Bray:1994}. In systems with short range
interactions, the initial droplets are of a typical radius which does
not grow with the system size. As a result, the characteristic
time for the formation of such droplets is also 
independent of the system size. On the other hand, in systems with
long range interactions the relaxation time scale diverges with the system 
size $N$. This may result 
 in long lived, quasistationary states (QSS) which
are not the true equilibrium state of the system but which relaxes 
to the equilibrium state on time scales that increase algebraically with $N$.

Such long-lived states have been seen, for instance, in numerical
studies of long-ranged spin model \cite{Yamaguchi:2004}. These QSS
have been explained as stable stationary states of the Vlasov
equation and may lead to anomalous diffusion
\cite{Bouchet:2005a,Bouchet:2005b}. We note that an alternative
explanation, both for the existence of QSS and for anomalous
diffusion has been proposed in the context of Tsallis non
extensive statistical mechanics \cite{Latora:2001,Pluchino:2004a}
(see \cite{Yamaguchi:2004} and \cite{Bouchet:2005a} for further
discussions). Recent studies have considered the possible
prediction of QSS using the equilibrium statistical mechanics of
the Vlasov equation
\cite{Barre:2004,Antoniazzi:2006,Chavanis:2006}. The issue of the
robustness of QSS when the Hamiltonian is perturbed by short range
interactions \cite{Campa:2006} or when the system is coupled to an
external bath \cite{Baldovin:2006a,Baldovin:2006b} has also been
addressed, and it was found that while the power law behavior
survives, the exponent may not be universal. Such a slow relaxation  is
not the only possible behavior in these systems \cite{note}. A
recent consideration of the thermodynamic stability of a  mean
field Ising model with stochastic dynamics has found the
relaxation time to be logarithmic in $N$ \cite{Mukamel:2005}. It
is thus of interest to study the slow relaxation processes in
systems with long range interactions and explore in more detail
the possible resulting time scales involved.

With this aim in mind, we consider the microcanonical dynamics of
a generalized Hamiltonian Mean Field (HMF) model which is a simple
prototype of long-ranged systems by adding a global anisotropy term or
an on-site potential energy term to the Hamiltonian. The basic HMF
model (\cite{Antoni:1995}, see also
\cite{Zaslavsky:1977,Konishi:1992,Inagaki:1993,Pichon:1994,Del-Castillo:2002}) describes a system
of $N$ classical XY rotors with mean field coupling. Adding new
terms to the Hamiltonian allows one to vary an external parameter
(such as anisotropy) and explore a richer phase diagram. Our
analysis of the Vlasov equation for these models shows that both
logarithmic and power law behavior are generically present in
long-ranged systems. It is found that at low energies the
non-magnetic solution is dynamically unstable and the system relaxes to the 
magnetically ordered state on a 
logarithmic time scale which follows from the dynamic instability
of Vlasov equation. At higher energies, but still within 
the magnetic state, the non-magnetic solution
becomes linearly stable (although it is not the true
equilibrium state of the system) and the relaxation takes place on
algebraically diverging time scales and QSS are observed. We show
the existence of QSS using analytic relations for the marginal
stability of the Vlasov equation. These results give further
insight into the robustness of QSS states when the interaction
potential is perturbed.

Most of the QSS studied so far have dealt with homogeneous situations
(namely states whose distribution functions do not depend on the
angle or spatial variable). From a theoretical point of view,
inhomogeneous QSS should exist in the same way as homogeneous
ones. The main reason why such states have not been studied in
detail is the difficulty to deal theoretically with the
inhomogeneous marginal stability equation. The QSS of the HMF
model with on-site potential described in Section \ref{onsite} is
an example of inhomogeneous QSS. Moreover, the new method we
propose to study them can be applied to other situations, and also
for instance to the usual isotropic HMF model.

The relaxation times mentioned above are interesting from another
viewpoint. Since the rationale for the existence of QSS is based
on the approximation of the $N$ particle dynamics by a Vlasov
dynamics, a crucial issue is the understanding of
the validity of such an approximation. In a classical work by Braun and Hepp 
\cite{Braun:1977}, it was proved that this approximation is valid for time
\( t \) smaller than \( t_{V}(N) \) 
for smooth interaction potential and large enough $N$. Following the 
reasoning of
\cite{Braun:1977} allows one to conclude that a lower bound for \(
t_{V}\left( N\right)  \) is proportional to \( \ln N \) for large
\( N \). A recent paper \cite{Caglioti:2005} showed that, for
initial conditions close to some homogeneous QSS, a lower bound
for \( t_{V}\left( N\right)  \) scales like \( N^{1/8} \). In
Section~\ref{validity}, we give two new results concerning this
issue. First, we prove that the Braun and Hepp result is actually
optimal. More precisely, we show that for some initial conditions,
the kinetic description is not valid for time larger than \( \sim
\ln N \). Second, we argue that most of the trajectories will have 
 \( t_{V} \) either equal to the life time of a QSS (possibly 
algebraic), or logarithmically long \( t_{V} \) depending on the way
the relaxation towards equilibrium takes place.

The rest of the paper is organised as follows. In Section~\ref{isotropic},
we define the HMF model and study its dynamically unstable state in detail
using Vlasov equation and numerical simulations. The anisotropic HMF model
is the subject of
Section~\ref{anisotropic} in which the dynamical phase diagram is obtained
analytically in the energy-anisotropy plane. The model with on-site
potential is introduced and discussed in Section~\ref{onsite}. The issue
of the time scale over which Vlasov equation holds is discussed
in Section~\ref{validity}. Finally,
we conclude with a summary and open questions in Section~\ref{conclusion}.

\section{Isotropic Hamiltonian Mean Field Model}
\label{isotropic}

In this section, we consider the dynamics of the Hamiltonian Mean
Field (HMF) model which is defined by the Hamiltonian 
\be
H=\sum_{i=1}^N \frac{p_{i}^{2}}{2}+\frac{1}{2N} \sum_{i,j=1}^N
\left[ 1-\cos(\theta_{i}-\theta_{j}) \right]~, 
\ee 
where $\theta_i$ and $p_i$ are the phase and momentum of the $i$th particle 
respectively, and $N$ is
the number of particles. In the equilibrium state, a second
order phase transition between the ferromagnetic and paramagnetic
state occurs at the critical energy density $\epsilon_c=3/4$. This
has been shown in the canonical ensemble \cite{Antoni:1995}
and later verified for microcanonical ensemble using large
deviations method \cite{Barre:2005}.

The time evolution of this system starting far from the equilibrium state has
been studied using Hamiltonian dynamics. The angle $\theta_i$ and momentum
$p_i$ of the $i$'th particle obey
\bea
\frac{d \theta_i}{dt} &=& p_i \label{theta}\\
\frac{d p_i}{dt} &=& -m_x \sin \theta_i+ m_y \cos \theta_i~, \label{mom}
\eea
where $m_x$ and $m_y$ are the components of the magnetization density
\be
\vec{m}= \left(\frac{1}{N}\sum_{i=1}^{N} \cos \theta_i,
\frac{1}{N}\sum_{i=1}^{N} \sin \theta_i \right)~.
\ee
The dynamics conserves the total energy and the total momentum.
We start with an initial condition with randomly distributed
$\theta_i \in [-\pi, \pi]$ so that the average magnetization is zero
and the standard deviation about the mean is of the order $\sim 1/\sqrt{N}$.
To fix the
total energy density $\epsilon$,
the momentum is chosen from a distribution $f^{(0)}(p)$ with $p_i$ lying
in the interval [$p_{\rm{min}}$, $p_{\rm{max}}$]. In this article, we 
consider the following choices of momentum distribution
\be
f^{(0)}(p)=\cases { (1/2 p_0) & {, $p \in [-p_0, p_0]$} \cr
\sqrt{\beta/2 \pi} \; {\rm exp} \left(-\beta p^2/2 \right) & {, $ p \in (-\infty,\infty)$}
}
\ee
where the parameter $p_0$ in the uniform (or waterbag) distribution and
$\beta$ in the Gaussian case are related to the energy as
\bea
p_0 &=& \sqrt{6 \epsilon-3}~, \nonumber \\
\beta &=& 1/(2 \epsilon-1) ~.
\eea
The equations of motion (\ref{theta}) and (\ref{mom}) are integrated using
symplectic fourth order integrator
with time step $dt=0.1$. The reference coordinate axes in which $\theta_i$ is
measured is specified by the initial magnetization.

To study the dynamical behavior of magnetization, we first recall the
classical computation of the Vlasov equation \cite{Nicholson:1983,Braun:1977,Spohn:1991,Firpo:1998,Elskens:2002} (see also \cite{Antoni:1995} for the HMF model).
The probability density $f_d(\theta, p,t)$ which counts the number of
particles
with angle $\theta$ and momentum $p$ at time $t$ can be written as
\be
f_d(\theta,p,t)=\frac{1}{N} \sum_{i=1}^{N} \delta(\theta_i(t)-\theta) \; \delta(p_i(t)-p)~. \label{fd}
\ee
Taking the time derivative of both sides of the above equation 
and using the canonical equations of
motion, one obtains
\be
\frac{\partial f_d}{\partial t}+p \frac{\partial f_d}{\partial \theta}-\frac{\partial V}{\partial \theta} \frac{\partial f_d}{\partial p}=0~,
\ee
where the average potential $V(\theta,t)$ is given by
\be
V(\theta,t)=\int_{p_{\rm{min}}}^{p_{\rm{max}}} dp'\int_{-\pi}^{\pi} d\theta' (1-\cos(\theta-\theta')) f_d(\theta',p',t)~.
\ee
Expanding (\ref{fd}) to leading orders in $1/\sqrt{N}$ (see below), we obtain
the Vlasov
equation obeyed by the (smooth) distribution $f(\theta,p,t)$ for infinite $N$,
\be
\frac{\partial f}{\partial t}+p \frac{\partial f}{\partial \theta}-\frac{\partial V}{\partial \theta} \frac{\partial f}{\partial p}=0~. \label{vlasov}
\ee
It is easily verified that the initial condition with angles
distributed uniformly and momentum chosen
from an arbitrary (normalised) distribution $f^{(0)}(p)$ is in fact
a stationary state of this equation. To deal with the finite $N$ case, we
treat the finiteness as
a perturbation about the homogeneous stationary state of the infinite system,
\be
f(\theta,p,t)=\frac{1}{2 \pi} f^{(0)} (p)+ \lambda f^{(1)} (\theta,p,t)~, \label{lambda}
\ee
 where, after linearization, the perturbed distribution $f^{(1)}$ is a solution of the
integro-differential equation 
\be 
\frac{\partial f^{(1)}}{\partial
t}+ p \frac{\partial f^{(1)}}{\partial \theta}- \frac{1}{2 \pi}
\frac{\partial f^{(0)}}{\partial p} \int dp' d \theta'
f^{(1)}(\theta',p',t) \sin(\theta-\theta')=0~. \label{vlasov_iso}
\ee 
Since the initial angles and momentum of the $N$ particles are
sampled according to the distribution $f^{(0)}$, the small
parameter $\lambda$ is of order $1/\sqrt{N}$.

We now study the linear dynamics about the distribution
$f^{(0)}(p)/2 \pi$ by considering the eigenmodes of (\ref{vlasov_iso}). A first treatment of the linear stability of stationary solutions of 
the HMF equation can be found in \cite{Inagaki:1993} 
(see also \cite{Del Castillo:2000} and \cite{Yamaguchi:2004}). As is 
explained for instance in textbooks on plasma physics \cite{Nicholson:1983}, 
any arbitrary function $f^{(1)}(\theta,p,t)$ cannot be 
decomposed into the eigenmodes of the linearized equation (\ref{vlasov_iso}). 
However when unstable modes 
exist, after a short time the largest of them dominates the dynamics and 
therefore at sufficiently long times, the temporal behavior can be 
described by the eigenmodes. 
We define the Fourier modes $f_k^{(1)}(p,\omega)$ (and the conjugate $f_{-k}^{(1)}$) as the eigenmodes of (\ref{vlasov_iso}) of type :
\be f^{(1)}(\theta,p,t)= f_k^{(1)}(p, \omega) e^{i (k \theta+ \omega t)} ~. \label{fourier}
\ee
Since the last term in the Vlasov equation (\ref{vlasov_iso})
involves only $e^{\pm i \theta}$, the Fourier modes must have $k=\pm1$. 
The coefficients $f_{\pm
1}^{(1)}$ are then determined by
\be f^{(1)}_{\pm 1}(p,\omega)+
\frac{1}{2} \frac{\partial f^{(0)}}{\partial p}
\frac{\int_{p_{\rm{min}}}^{p_{\rm{max}}} dp' f_{\pm
1}^{(1)}(p',\omega)}{p \pm \omega}=0~. \label{f1}
\ee
Integrating over $p$ on both sides, one gets
\be I_{\pm}
(1-J_{\pm})=0
\ee
where
\be
I_{\pm }= \int_{p_{\rm{min}}}^{p_{\rm{max}}} dp \; f_{\pm
1}^{(1)} (p) \;\;,\;\; J_{\pm}=-\frac{1}{2}
\int_{p_{\rm{min}}}^{p_{\rm{max}}} \frac{dp}{p \pm\omega}
\frac{\partial f^{(0)}}{\partial p}~. \l{I-J}
\ee
The frequency $\omega$ is thus found from the condition
$J_{\pm}=1$. For initial distributions $f^{(0)}$ which are even for the variable $p$, this condition yields an equation in $\omega^2$.
We now consider specific choices of momentum distribution
$f^{(0)}(p)$. For uniformly distributed initial momentum, the
frequency determined using the condition $J_{\pm}=1$ works out to
be \cite{Antoni:1995}
\be \omega^2=6 \left( \epsilon- \frac{7}{12} \right)~.
\label{omega2}
\ee
For $\epsilon > \epsilon^*=7/12$, unstable modes do not exist and the Vlasov 
equation is linearly
stable. It is however unstable for
$\epsilon < \epsilon^*$ and the perturbation $f^{(1)}(\theta,p,t)$ 
grows exponentially fast towards
the equilibrium state. Setting $\omega^2= -\Omega^2$ for $\Omega$
real, we have 
\bea
f^{(1)}(\theta,p,t) &=&  A f^{(1)}_1(\theta,p)e^{\Omega t} \label{f11}
\eea
where $A$ is a constant. As mentioned above, this time dependence is valid 
 for times $t >> 1/\Omega$. To treat the behavior at short times, the finite 
$N$ behavior in the initial condition must be taken into account 
\cite{Nicholson:1983}.

\begin{figure}
\begin{center}
\input{unstb2.tex}
\input{unstb1.tex}
\caption{(a) Time evolution of the average magnetization $m(t)$
in the unstable phase at $\epsilon=0.55$ for two values of $N$. 
(b) Data collapse for the scaled magnetization $\sqrt{N} 
m_x(t)$ vs. $t$ on the semi logarithmic  scale. The slope $\Omega=1/\sqrt{5}$ 
of the solid line is given by (\ref{omega2}). The data have been averaged
over $200$ histories for each $N$. } \label{unstb}
\end{center}
\end{figure}
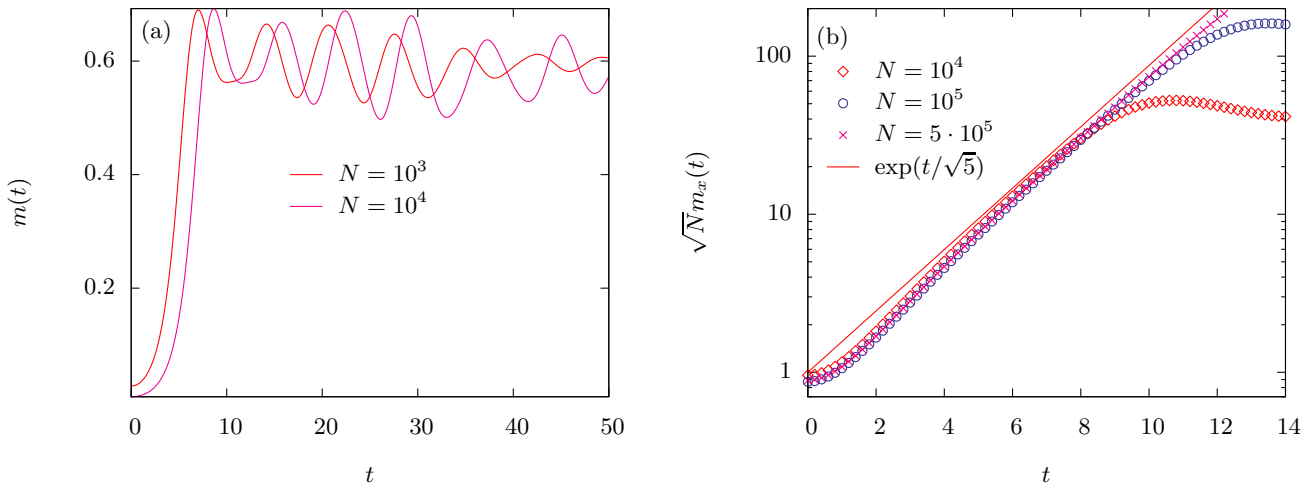

The average magnetization along the $x$ and $y$
axes, which are the observables of interest, can be written as
\bea
(m_x (t),m_y(t)) &=& \int_{p_{\rm{min}}}^{p_{\rm{max}}} dp \; \int_{-\pi}^{\pi} d \theta \; (\cos \theta, \sin \theta) \; f(\theta,p,t) \label{mvec_ex} \\
&=& \lambda \int_{p_{\rm{min}}}^{p_{\rm{max}}} dp \; \int_{-\pi}^{\pi} d \theta \; (\cos \theta, \sin \theta) \; f^{(1)}(\theta,p,t) + {\cal O}(\lambda^2)~.
\label{mvec}
\eea
The magnitude of the average magnetization is given by $m=\sqrt{m_x^2+m_y^2}$ 
and grows as 
\be
m \sim \frac{1}{\sqrt N} e^{\Omega t}~. \label{m_iso}
\ee
The results of our simulations in Fig.~\ref{unstb}a show that
after a transient, the magnetization grows
exponentially, as expected on the basis of preceding equation. Since the order 
of magnitude of the constant $A$ in 
(\ref{f11}) is proportional to $1/\sqrt{N}$, the 
scaled magnetization $\sqrt{N} m_x$ in Fig.~\ref{unstb}b collapses into a
single curve. The
growth rate is also in agreement with $\Omega$ obtained in (\ref{omega2}).  
The above perturbative
analysis cannot hold at long times as the linearization of the
Vlasov equation breaks down when the magnetization reaches a value of order 
one. A similar analysis can be carried out for Gaussian distributed initial
momentum. In this case, we obtain that the modes exist only for 
$\epsilon < \epsilon^*=3/4$ and the eigenvalues of the modes are given by 
(37) with $D=0$. Thus, (21) is obeyed in this case as well with the 
corresponding eigenfrequency. 
Using (\ref{m_iso}), we see that the time scale on which the system 
acquires a finite $m$ diverges as $\ln N$. 
Although
this is the same behavior as in the thermodynamically unstable
phase of Ising model \cite{Mukamel:2005}, the origin of
the logarithmic time scale here is dynamical while it follows from an 
argument based on thermodynamics in the Ising case.

So far we have discussed the dynamical instability.  The above argument
gives a $\ln N$ time scale for relaxation for $\epsilon < \epsilon^*$. Statistical mechanics predicts that homogeneous states are thermodynamically unstable for all
(allowed) values of $\epsilon < \epsilon_c=3/4$. For $\epsilon^* <
\epsilon < \epsilon_c$, the distribution $f^{(0)}(p)$ is dynamically stable but thermodynamically unstable. The relaxation is then not due to a dynamical instability and one then observes
a $N^{1.7}$ time scale (see \cite{Yamaguchi:2004}). 

\section{Anisotropic Hamiltonian Mean Field Model}
\label{anisotropic}

In this section, we consider the dynamics of the anisotropic HMF
model defined by the Hamiltonian
\be
H=\sum_{i=1}^N \frac{p_{i}^{2}}{2}+\frac{1}{2N} \sum_{i,j=1}^N \left[ 1-\cos(\theta_{i}-\theta_{j})\right]- \frac{D}{2 N} \left[ \sum_{i=1}^N \cos \theta_{i} \right]^2
\label{H_an}~,
\ee
where the last term represents the energy due to a global anisotropy in
the magnetization along the $x$-axis. At zero temperature, if $D$
is positive, the equilibrium magnetization is along the $x$-axis.
Similarly for $D < 0$, the internal energy is lowered when the
magnetization is along the $y$ direction. For simplicity we
consider below the case $D>0$. To study the equilibrium phase
diagram, consider the partition function in the canonical
ensemble,
\be Z = \int \prod_{i=1}^N dp_i d \theta_i e^{-\beta H}
= \left(\frac{2 \pi e^{-\beta}}{\beta} \right)^{N/2} \int \prod_i
d \theta_i \; \mathrm{exp} \left[ \frac{\beta N}{2} \left((1+D)
m_x^2+ m_y^2 \right) \right]~.
\ee
On using a Hubbard-Stratonovich transformation \cite{Antoni:1995},
the integrals over angles can be rewritten as
\be
Z_\theta=
\frac{N}{2 \pi \beta \sqrt{(1+D)}} \int db_{x} \;db_y
\;\mathrm{exp} \left[ -N \left( \frac{b_y^2}{2
\beta}+\frac{b_x^2}{2 \beta (1+D)}-\ln \int_{-\pi}^{\pi} d \theta
\; e^{b_x \cos \theta+b_y \sin \theta} \right) \right]~,
\ee
where the double integral can be evaluated using the saddle point
method for large $N$, leading to the free energy per particle
\be
f=\frac{-\ln Z}{N \beta}=\frac{-1}{2 \beta} \ln \left( \frac{2
\pi}{\beta}\right)+\frac{1}{\beta} \left[
\frac{\overline{b}_x^2}{2 \beta (1+D)}+\frac{\overline{b}_y^2}{2
\beta }-\ln \int_{-\pi}^{\pi} d \theta \; e^{\overline{b}_x \cos
\theta+\overline{b}_y \sin \theta} \right].
\ee
In the above expression, $\overline{b}_x$ and $\overline{b}_y$ are 
determined by maximizing $Z_\theta$ with
respect to $b_x$ and $b_y$,  and are related to the equilibrium magnetization 
$\overline{m}_x$ and $\overline{m}_y$ 
along the $x$ and $y$ axis respectively as 
$\overline{b}_x=\beta (1+D)\overline{m}_x$ 
and $\overline{b}_y=\beta \overline{m}_y$. 
As explained above, for $D > 0$, the system 
orders along the $x$-axis and the magnetization $\overline{m}_x$ is 
determined by 
\bea
\overline{m}_x &=& \frac{\int_{-\pi}^{\pi} d \theta ~\cos \theta ~ e^{\beta (1+D)\overline{m}_x \cos \theta}}{\int_{-\pi}^{\pi} d \theta ~e^{\beta (1+D)\overline{m}_x \cos \theta}}~. 
\eea
Close to the critical point, the above transcendental equation 
can be expanded in a Taylor series about zero magnetization and we
obtain 
$\overline{m}_x^2= (8 \beta (1+D) -16)/((4-\beta (1+D)) \beta^2 (1+D)^2)$ for 
$D > 0$. The
inverse critical temperature at which magnetization vanishes is
given by $\beta_c=2/(1+D)$. The critical energy $\epsilon_c$ can
be calculated using the free energy expression above and we obtain
\be \epsilon_c= \left[ \frac{\partial (\beta f) }{\partial \beta}
\right]_{\beta=\beta_c}= \frac{3+D}{4} \;.
\ee
\begin{figure}
\psfrag{D}{$D$}
\psfrag{E}{$\epsilon$}
\psfrag{4E-3}{$4 \epsilon-3$}
\psfrag{C2}{$3/4$}
\psfrag{12E-7}{$12 \epsilon-7$}
\psfrag{C1}{$7/12$}
\psfrag{8E-6}{$8 \epsilon-6$}
\psfrag{mx0my0}{$\overline{m}_x,\overline{m}_y=0$}
\psfrag{mx}{$\overline{m}_x \neq 0$}
\psfrag{my}{$\overline{m}_y \neq 0$}
\psfrag{(a)}{(a)}
\psfrag{(b)}{(b)}
\includegraphics[width=0.445 \linewidth,angle=0]{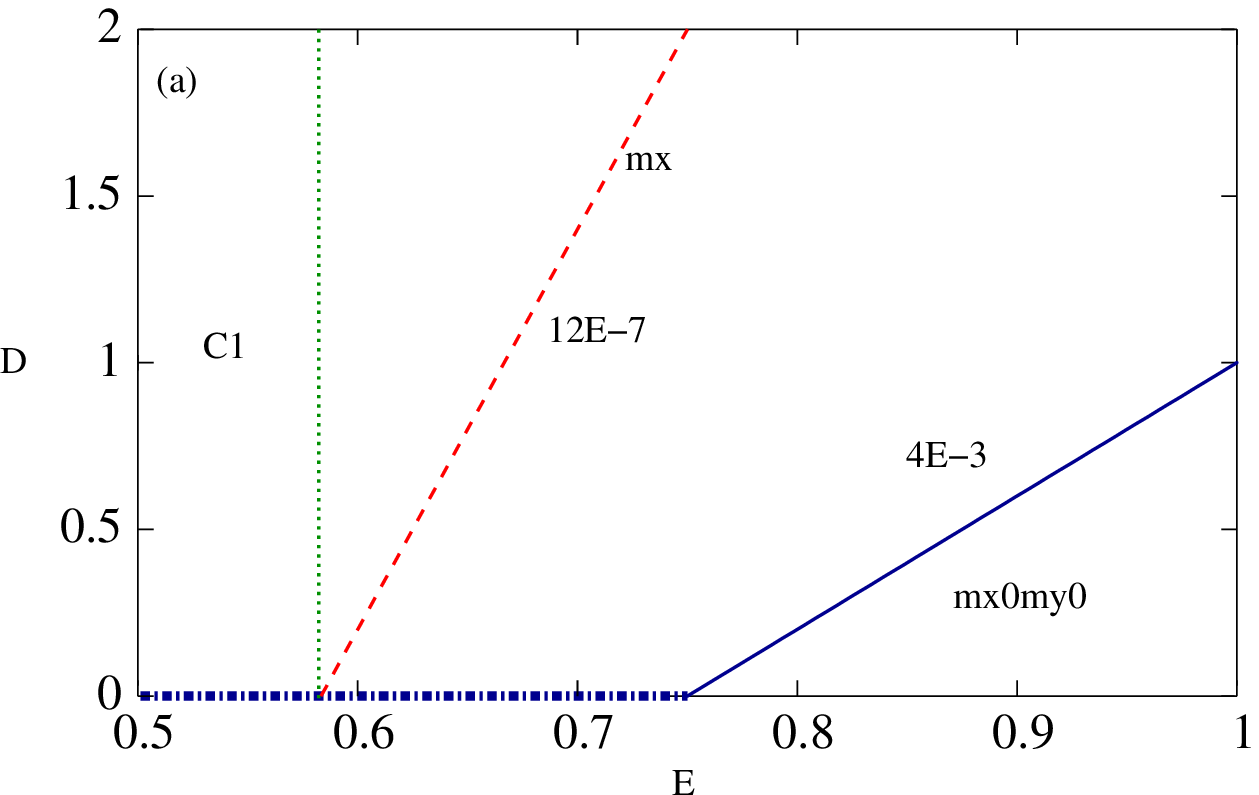}
\includegraphics[width=0.45 \linewidth,angle=0]{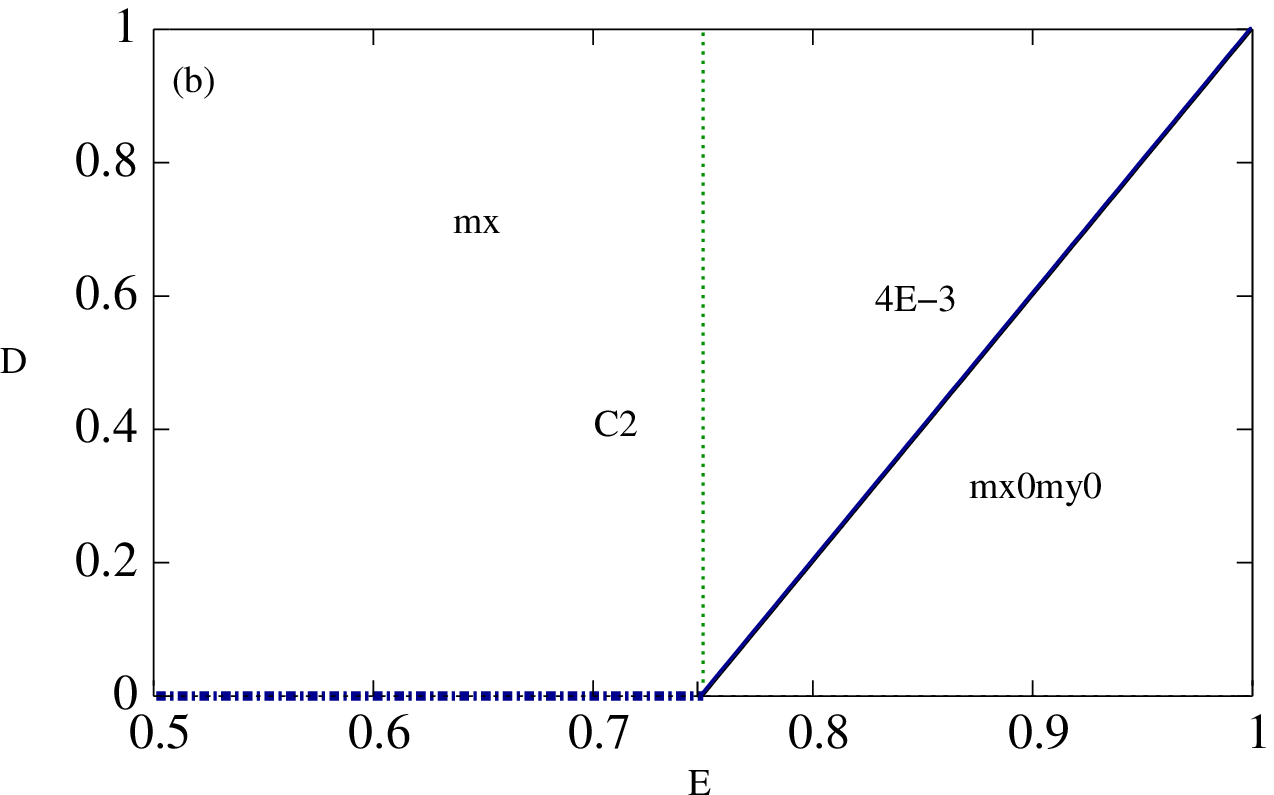}
\caption{Dynamical phase diagram in the $(\epsilon, D)$
plane for the initial momentum distribution chosen as
Waterbag (a) and
Gaussian (b). The solid line is the thermodynamic phase boundary
$D(\epsilon_c)$ and the broken line is the dynamical boundary
$D(\epsilon^*)$. The dotted line gives the line where $\omega '$
vanishes. The dashed dotted  line is the first order line where
the magnetization changes its direction. Note that in the
Gaussian case, the dynamical and the thermodynamical boundaries coincide.}
\label{dyn_an}
\end{figure}

For the anisotropic Hamiltonian, the equations of motion are
similar to those for the isotropic case, (\ref{theta}) and
(\ref{mom}), except that $m_x$ is replaced by $(1+D) m_x$. The
average potential appearing in the corresponding Vlasov equation
(\ref{vlasov}) is now given by
\be V(\theta,t)=\int_{p_{\rm
min}}^{p_{\rm max}} dp' \int_{-\pi}^{\pi} d\theta' \left[
1-\cos(\theta-\theta') -D \cos \theta \cos \theta' \right]
f(\theta',p',t)~.
\ee
It can be checked that the homogeneous state (in $\theta$) is a
stationary state of the Vlasov equation for the anisotropic HMF
model so that the distribution $f(\theta,p,t)$ can be written as
(\ref{lambda}). The distribution $f^{(1)}$ is now a solution of
the following equation, \be \frac{\partial f^{(1)}}{\partial t}+ p
\frac{\partial f^{(1)}}{\partial \theta}- \frac{1}{2 \pi}
\frac{\partial f^{(0)}}{\partial p} \int dp' d \theta'
f^{(1)}(\theta',p',t) \left[ \sin(\theta-\theta')+D \sin \theta
\cos \theta'\right]=0~. \ee As before, going to the Fourier space
and picking the coefficient of $e^{\pm i \theta}$, we obtain \be
I_{\pm }=  \left[ \left(1+\frac{D}{2} \right) I_{\pm }
+\frac{D}{2} I_{\mp } \right]  J_{\pm} \ee where $I_{\pm}$ and
$J_{\pm}$ are given in (\ref{I-J}). For $D \neq 0$, the frequency
$\omega$ is then determined through \be (1+D) J_+ J_-
-\left(1+\frac{D}{2}\right) (J_+ + J_-)+1=0 \label{omega_an} \ee
which is a {\it bilinear} equation unlike in the isotropic case.
We now find the dynamical phase diagram and the frequency in the
unstable phase for two choices of initial momentum distribution.

\begin{figure}
\begin{center}
\input{wU.tex}
\input{wS.tex}
\caption{Temporal behavior of the magnetization in the anisotropic HMF model 
for uniformly distributed initial momentum. (a) 
The three curves in the unstable phase show scaled
magnetization $\sqrt{N} m_x(t)$ for $\epsilon=0.55$ and $D=0.9$.
The slope of the solid lines
is given by $\omega$ in (\ref{omegap_w}). (b) Evolution in the stable phase 
with
$\epsilon=0.8$ and $D=0.9$ (top to bottom) for $N=500 (100), 2000
(50), 5000 (50), 10000 (5)$. The number of histories over which
data are averaged is given in parenthesis. The scaled data is
consistent with $N^{1.7}$ scaling of the quasistationary life
time.} \label{w_an}
\end{center}
\end{figure}
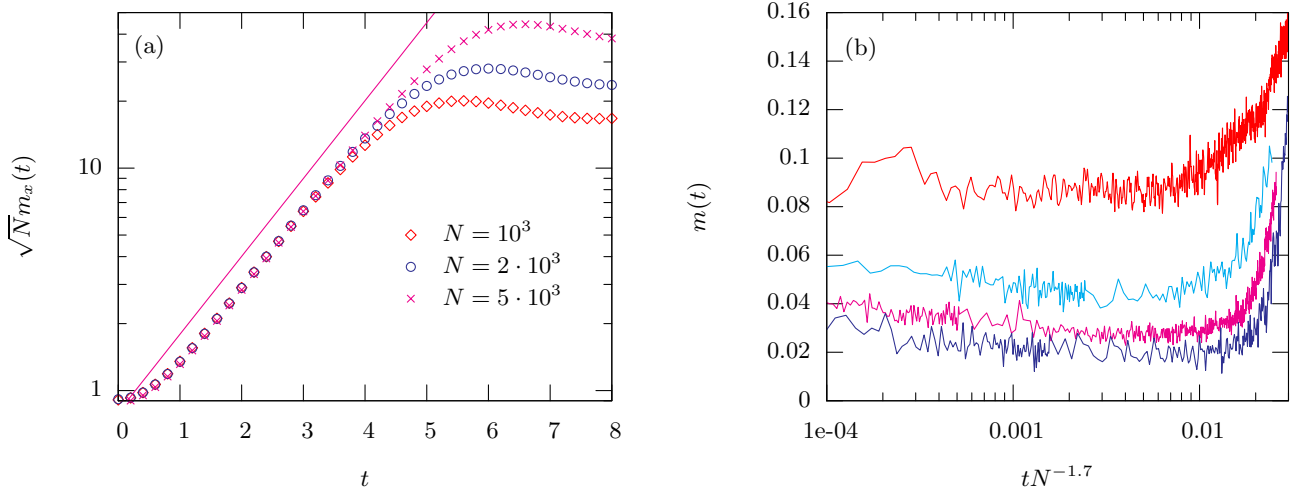

\noindent{\it Uniform distribution.-} For uniformly distributed
$f^{(0)}(p)$, the integrals $J_\pm$ appearing in (\ref{omega_an})
can be readily done and we obtain a fourth order equation for the
frequency $\omega$,
\be 4 (\omega^2-p_0^2)^2
+2~(2+D)~(\omega^2-p_0^2)+(1+D)=0
\ee
with the following four solutions
\bea
\omega^2&=& 6 \left(\epsilon- \frac{7+D}{12}  \right) \l{omegap_w}\\
{\omega{'}}^2 &=& 6 \left(\epsilon- \frac{7}{12} \right)
\l{omegan_w}~.
\eea

The frequency $\omega^2$ vanishes at $\epsilon^*=(7+D)/12$ while
${\omega '}^2$ becomes zero at $\epsilon '^* =7/12$. Thus in the
magnetically ordered phase, $\epsilon < \epsilon_c$, three regions
can be identified: a linearly stable region $\epsilon^* < \epsilon
< \epsilon_c$ where long relaxation time scales increasing as a
power of $N$ are expected, an unstable regime
$\epsilon'^* < \epsilon < \epsilon^*$ with a single mode
of instability namely $\Omega$, and another
unstable regime $\epsilon <\epsilon'^*$ with two modes of instability
$\Omega$ and $\Omega'$ where $\Omega^2=-\omega^2$ and $\Omega'^2=-\omega'^2$.  
Since $\Omega > \Omega'$, the magnetization
increases exponentially fast with rate $\Omega$.
Thus, in the last two regimes, the relaxation time scales
diverging as $\ln N$ are expected. The $(\epsilon, D)$ phase
diagram resulting from this analysis is given in Fig.
~\ref{dyn_an}a.

Our numerical results for the time evolution of the magnetization
in the unstable and the stable phases are shown in
Fig.~\ref{w_an}. For $\epsilon < \epsilon^*$ (Fig.~\ref{w_an}a),
data collapse of the curves for various system sizes is
observed when the magnetization is scaled with a factor $\sqrt{N}$
as in the last section. Thus, we again obtain $\ln N$ scaling for the
relaxation time. The growth rate in the unstable phase is also in
agreement with $\omega$ in (\ref{omegap_w}). For $\epsilon >
\epsilon^*$(Fig.~\ref{w_an}b), the magnetization stays close to its initial 
value
$\sim 1/\sqrt{N}$ for a long time which is consistent with
$N^{1.7}$ scaling as for the basic HMF model.

\noindent{\it Gaussian distribution.-}
For initial momentum chosen from Gaussian distribution, the dynamics
are always unstable and the frequency $\omega^2=-\Omega^2$, $\Omega$ real.
Consider the integrals $J_\pm$ defined in (\ref{I-J}),
\be
J_+=J_-=\frac{\beta}{2} \left[1- \Omega^2 \sqrt{\frac{\beta}{2 \pi}} \int_{-\infty}^{\infty} dp \;\frac{e^{-\beta p^2/2}}{p^2+\Omega^2} \right]
\ee
where we have used that the derivative of $f^{(0)}$ is an odd
function. The integral in the last term can be evaluated using the
Schwinger trick,
\be
\int_{-\infty}^{\infty}
dp \;\frac{e^{-\beta p^2/2}}{p^2+\Omega^2}
=\int_{-\infty}^{\infty} dp ~e^{-\beta p^2/2} \int_0^\infty dq
~e^{-q(p^2+\Omega^2)}= \frac{\pi}{\Omega} ~ \rm{exp}
\left(\frac{\beta \Omega^2}{2} \right) ~\rm{Erfc}
\left(\sqrt{\frac{\beta}{2}} \Omega\right)~.
\ee

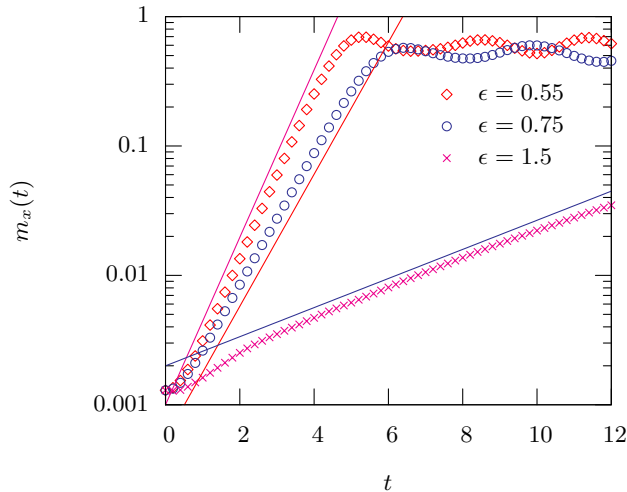
\begin{figure}
\input{gUp.tex}
\caption{Semi-log plot of the magnetization as a function of time 
for Gaussian $f^{(0)}(p)$
in the dynamically unstable, ferromagnetic phase for $N=5 \times 10^5, D=4$ and
various $\epsilon$. The lines have a slope $\Omega$
given by (\ref{Omegap}).}
\label{gU_an}
\end{figure}

Since the dispersion relation (\ref{omega_an}) is quadratic in
$J_+$  it has two solutions namely $J_+=1/(1+D)$ and $J_+=1$ with
respective frequencies $\Omega$ and $\Omega '$ which obey
\bea
1- \sqrt{\frac{\beta \pi}{2}} ~\Omega ~\rm{exp} \left(\frac{\beta \Omega^2}{2} \right) ~\rm{Erfc} \left(\sqrt{\frac{\beta}{2}} \Omega \right) &=& \frac{2}{\beta (1+D)} \l{Omegap} \\
1- \sqrt{\frac{\beta \pi}{2}} ~\Omega ' ~\rm{exp} \left(\frac{\beta \Omega '^2}{2} \right) ~\rm{Erfc} \left(\sqrt{\frac{\beta}{2}} \Omega ' \right) &=& \frac{2}{\beta} \l{Omegan}~.
\eea
The real frequencies $\Omega, \Omega'$ vanish at
$\epsilon^*=(3+D)/4$ and $\epsilon '^*=3/4$ respectively. The critical energy
$\epsilon^* (> \epsilon '^*)$ coincides
with $\epsilon_c$ as there is no stable phase when
the initial momentum is distributed according to a Gaussian
distribution. One is thus left with two unstable regimes, one with
a single unstable mode $(\epsilon'^* < \epsilon < \epsilon_c)$,
and the other with two unstable modes $(\epsilon < \epsilon'^*)$.
Since the left hand side of the above equations for $\Omega$ and
$\Omega'$ is a monotonically decreasing function lying between $1$
and $0$, one has $\Omega > \Omega '$ for $D>0$. Thus we expect the
growth rate of $m_x$ to be $\Omega$ for all energies below
$\epsilon_c$. The dynamical phase diagram corresponding to the
case of Gaussian initial distribution is given in 
Fig.~\ref{dyn_an}b. Our numerical results for the evolution of the
magnetization verifying the conclusions presented above are shown
in Fig.~\ref{gU_an}.

\section{Hamiltonian Mean Field Model with on-site potential}
\label{onsite}

We next consider the HMF model with cosine on-site potential whose
Hamiltonian is 
\be
H=\frac{1}{2}\sum_{i=1}^{N}p_{i}^{2}+\frac{1}{2N}\sum_{i,j=1}^{N}(1-\cos(\theta_{i}-\theta_{j}))+ W \sum_{i=1}^{N}\cos^{2} \theta_{i}~. \label{H_on} 
\ee 
where the last term in the Hamiltonian gives the energy due to an on-site 
potential. For positive (negative) $W$, the steady state magnetization is 
along the $y(x)$-axis and in the following, we assume $W > 0$.
The equilibrium properties of this model can be calculated following the 
same procedure outlined in Section \ref{isotropic}. The equilibrium 
magnetization $\overline{m}_y$ along the $y$ axis is 
 determined by the following equation,
\bea
\overline{m}_y &=& \frac{\int_{-\pi}^{\pi} d \theta ~\sin \theta ~ e^{\beta \overline{m}_y \sin \theta-\beta W \cos^2 \theta}}{\int_{-\pi}^{\pi} d \theta ~e^{\beta \overline{m}_y \sin \theta-\beta W \cos^2 \theta}}~. 
\eea
The magnetization vanishes at a critical temperature 
determined via 
$\beta_c^{-1}=(I_0(z)+I_1(z))/2 I_0(z)$ where $I_0$ and $I_1$ are the 
 the modified 
Bessel functions of the first kind and the argument $z=\beta_c W/2$. 
For positive $W \ll 1$, 
the critical temperature at which $m$ becomes zero is 
given as $(2+W)/4$ to leading order in $W$. The critical energy $\epsilon_c$ 
is then given by 
\be
\epsilon_c = \frac{1}{2 \beta_c}+\frac{1}{2}+\frac{W}{\beta_c} \approx 
\frac{3}{4}+\frac{5 W}{8}~. \l{on_crit}
\ee

For the model defined by (\ref{H_on}), the
canonical equations of motion  reads, 
\be
\frac{d\theta_{i}}{dt}=p_{i},\quad\frac{dp_{i}}{dt}=-m_x
\sin\theta_{i}+m _y\cos\theta_{i}+2 W
\cos\theta_{i}\sin\theta_{i}~.\label{eq:canonical} 
\ee 
As in the
previous sections, for a large number of particles the dynamics is
well approximated by the Vlasov equation (\ref{vlasov}) with the
following potential
\be V[f]=-\int_{p_{\rm{min}}}^{p_{\rm{max}}} dp'\int_{-\pi}^{\pi}
d\theta' \cos(\theta-\theta') f(\theta',p',t)+ W \cos^{2}\theta~.
\ee
Unlike for the models considered in the preceding sections, the
homogeneous state is no longer a stationary state of the Vlasov
equation  due to the presence of the last term in the potential
$V$. However, the distribution function $f(\theta,p) =
\Phi(e(\theta,p))$, with arbitrary function $\Phi$, where the
single particle energy $e$ is given by
\be
e(\theta,p)=\frac{p^{2}}{2}+V\left(\theta\right)=\frac{p^{2}}{2}-m_x\cos\theta-m_y\sin\theta+W\cos^{2}
\theta~. \label{eq:individual-energy}
\ee
are stationary solutions of the Vlasov equation  (\ref{vlasov}).
This can be easily seen by differentiating Vlasov equation
(\ref{vlasov}) and by noting that the potential energy $V$ is a
function of the angle variable only. We stress that the
magnetization values $m_x$ and $m_y$ must be self consistently
determined. The particular case $f^{(0)}\left(\theta,p\right) \sim
\exp\left(-\beta e\left(\theta,p\right)\right)$ is the statistical
equilibrium density.

A special class of
 stationary distributions is given by
$f^{(0)}(\theta,p)=\Phi(p^2/2+W \cos^2 \theta)$. The fact that
$m_x=m_y=0$ follows by symmetry arguments. In the following, we
will study a waterbag stationary state distribution i.e. $\Phi$ is
a step function \be f^{(0)}(\theta,p)=\cases {A  & {,
$\frac{p^2}{2}+W\cos^2 \theta<E$} \cr
                          0 &  {, $\rm{otherwise}~.$}
}
\ee
The distribution function $f^{(0)}(p)$ is thus constant over the domain
${\cal {D}}$ defined by
\be
\left|p\right|<p_{0}\left(\theta\right)=\sqrt{2\left(E-W\cos^{2}\theta\right)}\label{eq:p0}~,
\ee
and is simply connected for $W < E$. The study of the case $W > E$ can be 
done
following ideas similar to those described below. The normalization constant
$A$ is determined using
$\int_{-\pi }^{ \pi}d\theta\int_{-p_0(\theta)}^{p_0(\theta)} dp\, f^{(0)}(\theta,p)=1$ and we have
\be
\frac{1}{2A}=\int_{-\pi}^{\pi}d\theta\,\sqrt{2\left(E-W\cos^{2}\theta\right)}\label{eq:A}~.
\ee
The parameter $E$ can be related to the conserved initial energy $\epsilon$
by performing the integration over the $p$ variable in (\ref{H_on}) and
we obtain
\be
\epsilon= A\int_{0}^{2\pi}d\theta\,\left\{ \frac{1}{3}\left[2\left(E-W\cos^{2}\theta\right)\right]^{3/2}+2W\cos^{2}\theta\left[2\left(E-W\cos^{2}\theta\right)\right]^{1/2}\right\} \label{eq:H_E_D}~.
\ee

To compute the linear stability threshold, as before, we linearize
the dynamics close to the stationary solution, $f=f^{(0)}+\lambda
f^{(1)} \exp (i\omega t )$. The perturbation $f^{(1)}$ then
satisfies
\be i\omega f^{(1)}+p\frac{\partial f^{(1)}}{\partial
\theta}-\frac{dV[f^{(0)}]}{d\theta} \frac{\partial
f^{(1)}}{\partial p}-\frac{dV[f^{(1)}]}{d\theta} \frac{\partial
f^{(0)}}{\partial p}=0 \label{eq:Vlasov_Linearisee}~.
\ee
We will
determine the energy $\epsilon^*$ which corresponds to the neutral
mode $\omega=0$. Below this energy, the dynamics is expected to
be unstable and stable above it. At $\omega=0$, the above equation
can be written explicitly as
\be p \frac{\partial
f^{(1)}}{\partial \theta}- A \left[ \delta(p+p_0
(\theta))-\delta(p-p_0 (\theta)) \right] \int dp' d \theta' \sin
(\theta-\theta') f^{(1)}(\theta', p',\omega) + 2 W \sin \theta
\cos \theta \frac{\partial f^{(1)}}{\partial p}=0~.
\ee
We solve this last equation by a formal expansion in terms of
Dirac distributions $\delta$ and its order $n$ derivatives
$\delta^{(n)}$ :\begin{equation}
f^{(1)}=\sum_{n=0}^{\infty}a_{n}\left(\theta\right)\delta^{(n)}\left(p+p_{0}\left(\theta\right)\right)+b_{n}\left(\theta\right)\delta^{(n)}\left(p-p_{0}\left(\theta\right)\right)\label{eq:Ansatz}\end{equation}
The equations for $a_{n}$ and $b_{n}$ and their solutions can be
found recursively. We report the analysis only for $a_{0}$ and
$b_{0}$ :
\begin{equation}
\frac{da_{0}}{d\theta}=\frac{db_{0}}{d\theta}=A\frac{\sin\theta
m_{x}\left[\delta f\right]-\cos\theta m_{y}\left[\delta
f\right]}{p_{0}\left(\theta\right)}\label{eq:a0_b0}
\end{equation}
The magnetization 
$m_{x}\left[\delta f\right]$ must be determined self-consistently
from the distribution function. Using (\ref{eq:Ansatz}), the formula 
(\ref{eq:a0_b0}) and (\ref{eq:A}),
we obtain
\be
\int_{-\pi}^{\pi}d\theta\,\sqrt{2\left(E-W\cos^{2}\theta\right)}=\int_{-\pi}^{\pi}d\theta\,\frac{\sin^{2}\theta}{\sqrt{2\left(E-W\cos^{2}\theta\right)}}\label{eq:marginal_1}~.
\end{equation}
This is the equation for the marginal stability of the inhomogeneous
waterbag distribution function. We define
\be
I(x)=\frac{1}{\int_{0}^{2\pi}d\theta\,\sqrt{1-x\cos^{2}\theta}} \int_{0}^{2\pi}d\theta\,\frac{\sin^{2}\theta}{\sqrt{1-x\cos^{2}\theta}}
~
\ee
which can be expressed in terms of complete elliptic functions
of the first and second type. The equation for the marginal stability
for inhomogeneous Water Bag distributions then reads
\begin{equation}
2E=I\left(\frac{W}{E}\right)\label{eq:marginal}~.
\end{equation}
One can prove that $I$ is a strictly decreasing function from the
interval $[0;1]$ onto the interval $[1/2;1]$. $I^{-1}$is thus an
increasing function from $[1/2;1]$ onto $[0;1]$. From this, one
can prove that, the equation $I^{-1}\left(2E\right)=W/E$ has a single
solution $E^*\left(W\right)$, for each value of $W$ in the range
$[0;1/2]$, and no solutions for $W>1/2$. For $W=1/2,$ we have $E^*=W$
; this is the limit above which the inhomogeneous Water-Bag cease
to be simply connected (see the discussion below  (\ref{eq:p0})).
For values larger than $W=1/2$, the transition value $E^*\left(W\right)$
could be studied by considering doubly connected domains.

\begin{figure}
\begin{center}
\includegraphics[height=7cm,width=8cm]{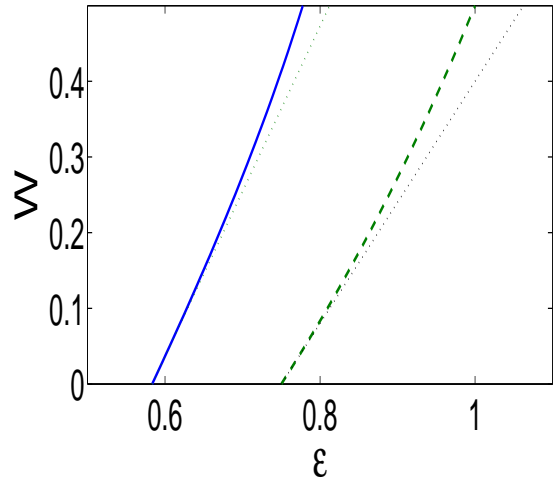}
\caption{Dynamical phase diagram in the $(\epsilon, W)$ plane for
the inhomogeneous uniform distribution of the HMF model with
on-site potential.
The bold line is the line of marginal stability $W(\epsilon^*)$ and the 
thermodynamic phase boundary $W(\epsilon_c)$ is shown as a 
dashed line. The dotted lines are the respective curves obtained within linear
approximation for $W \ll 1$ given in (\ref{on_crit}) and (\ref{on_stab}).} 
\label{fig:marginal_stability}
\end{center}
\end{figure}

For $W<<1$,
equation (\ref{eq:marginal_1}) can be easily linearized and we
obtain $E^*\left(W\right)=1/4+3W/8+O\left(W^{2}\right)$.
Now using (\ref{eq:H_E_D}), we can compute the critical  $\epsilon^*\left(W\right)$ as a function of $W$. For instance, for
$W<<1$, we obtain
\be
\epsilon^*\left(W\right)=\frac{7}{12}+\frac{11W}{24}+O\left(W^{2}\right)~. 
\l{on_stab}
\ee
For $W=0$,
we obtain the critical energy for the homogeneous Water Bag
distribution $\epsilon^*=7/12$ (see \cite{Yamaguchi:2004}). 
For any value $0\leq W\leq1/2$, the critical value of the energy can easily be computed numerically from (\ref{eq:marginal}). 
Figure \ref{fig:marginal_stability} shows the curve of marginal
stability.

At energies below $\epsilon^*(W)$, where the waterbag state is 
unstable, the relaxation time grows logarithmically with the system size
as in the case of the model with global anisotropy. This is a result of the 
fact that the initial magnetization of the waterbag state is of order 
$1/\sqrt N$. For energies above  $\epsilon^*(W)$ where the waterbag state is 
linearly stable, we expect a longer relaxation time, which grows 
algebraically with $N$.


\section{How long is the Vlasov approximation valid ?}
\l{validity}

In this section, we discuss the validity of the approximation of the
$N$-particle phase-space distribution
\be f_d(\theta,p,t)=\frac{1}{N} \sum_{i=1}^{N}
\delta(\theta_i(t)-\theta) \; \delta(p_i(t)-p) \ee
by a smooth density function $f_s(\theta,p,t)$, where both $f_d$ and 
$f_s$ satisfy the
Vlasov equation (\ref{vlasov}) (mean field approximation). We ask:
how long is this approximation justified?

We consider an ensemble of initial conditions for the $N$-particles  
$\left\{ (\theta_i(t=0),p_i(t=0))_{1\leq i \leq N} \right\}$. We suppose that
 the corresponding $f_d(t=0)$ is close to some smooth distribution function 
$f_s(t=0)$ (i.e. the distance between 
$f_d$ and $f_s$ converges to zero as $N$ goes to infinity). If for almost 
all initial conditions, $f_d$ remains close to the smooth $f_s$, we 
say that 
the system has a kinetic behavior: all trajectories remain close to each 
other. 
The kinetic evolution is then the solution of the Vlasov equation with initial 
condition $f_s(t=0)$. The issue of the validity of the mean field 
approximation, or equivalently of the validity of the kinetic description, is 
to know for how long all trajectories remain close to $f_s$. Since 
$f_d$ satisfies
 the Vlasov equation, this issue is related to the stability of the 
Vlasov equation.

Let us first summarize the known results.
The validity of this mean field approximation for large $N$ has
been established mathematically, for smooth potential $V$, by Braun
and Hepp~\cite{Braun:1977} (see also~\cite{Spohn:1991}). More
precisely, the theorem of Braun and Hepp states that for a mean-field
microscopic two-body smooth potential, the distance between two
initially close solutions of the Vlasov equation increases at
most exponentially in time. Indeed, if $f(t)$ and $g(t)$ are two
solutions, if $\Delta(f,g)(t=0)$ is sufficiently small, then $\Delta(f,g)(t)
\leq \Delta(f,g)(0)\exp(at)$, where $\Delta$ is the Wasserstein distance and
$a$ is a constant.

This result can be applied to the approximation of the
$N$-particle Hamiltonian dynamics by the Vlasov equation. We
consider, for the $N$ particle dynamics, an ensemble of initial
conditions $\left\{ (\theta_i,p_i)_{1\leq i \leq N} \right\}$
distributed according to the measure
$f(\theta_1,p_1,...,\theta_N,p_N)=\Pi_{i=1}^{N}f_s(\theta_i,p_i)$.
For large $N$, for a typical initial condition, the phase-space
initial distribution $f_d(\theta,p,0)$ will be close to $f_s(\theta,p,0)$.
Typically $\Delta(f_d,f_s)(t=0) = \mathcal{O}(1/N^{\alpha})$ with 
$\alpha > 0$. Let us
consider $f_s$ the solution of the Vlasov equation with initial
condition $f^{(0)}$. Because $f_d$ and $f_s$ are both solutions for the
Vlasov equation, we can apply the Braun and Hepp's result. If we
define $t_V$ to be the time at which the error $\Delta(f_d,f_s)$ is of order 
unity, the theorem then implies that $t_V$ increases at least
as $\ln N$ when $N \to \infty$ (for this argument, 
see also~\cite{Yamaguchi:2004}).

In Sections~\ref{isotropic} and \ref{anisotropic}, we have
considered the special case where the initial distribution $f^{(0)}$ is a
stationary solution of the Vlasov equation. When $f^{(0)}$ is unstable, the 
perturbation
$f^{(1)}$ grows exponentially, we have explained and illustrated
that $t_V$ is proportional to $\ln N$. Thus the trajectories diverge from 
$f_s$ on a time scale given by $\ln N/\Omega$. This thus proves that the
Braun and Hepp's result for $t_V$ is not only a lower bound, but is
actually achieved.  After this time scale, the trajectories diverge and the 
system does not have a kinetic behavior any more.  

When $f^{(0)}$
is a {\it stable} stationary solution of the Vlasov equation, $t_V$ is the 
stability time of the
Quasi-Stationary State. A very recent work~\cite{Caglioti:2005}
has proven that the $N$-particle dynamics actually remains close
to the stationary solutions, at least for times of order
$N^{1/8}$, when the potential $V$ is sufficiently smooth. On
physical grounds, using kinetic theory, one expects the  validity
time to be of order $N$ for systems in which each particle is characterised by 
more than one dynamical variable. When no
resonance between trajectories is possible as is the case for 
systems with one dynamical variable, one expects validity times to be much 
larger than $N$~\cite{Bouchet:2005a}. This peculiarity of $1d$ systems has
been numerically observed in the HMF model, where times of order
$N^{1.7}$ have been measured for homogeneous Quasi-Stationary
states~\cite{Yamaguchi:2004}. This time scale seems to be
robust with respect to the perturbations of the Hamiltonian as we
have demonstrated for the anisotropic case in
Section~\ref{anisotropic}.

Let us now consider the more general case when the initial
condition is close to a distribution $f$ which is not stationary.
In accordance with the observed phenomenology for the Vlasov
equation, one expects that $f$ will have a rapid relaxation in a
finite time either towards a Quasi-Stationary state, or towards a
periodic solution, or towards a statistical equilibrium for the
Vlasov equation. Because this first stage takes place in times
which are of order one (which do not depend on $N$), one expects
that this initial relaxation will have a negligible effect on the
long time error. The validity time for the Vlasov approximation
will then be given by  the validity time of the Quasi-Stationary
states. Then times $t_V$ of order $N^\alpha$ are expected. We thus
conjecture that, for generic initial distributions, the
approximation by the Vlasov equation is valid over times which are
the life time for the Quasi-Stationary states.


\section{Conclusions}
\label{conclusion}

In this article, we studied the short time dynamics of models with
long-ranged Hamiltonians. In each case, starting from the initial
magnetization of order $1/\sqrt{N}$, the time required to achieve
a finite value of $m$ scales with the number $N$ of particles.
This behavior is different from that of the corresponding
short-ranged Hamiltonians where such time is of order unity. The
dynamics were studied by a numerical integration of the Hamilton's
equations of motion and a stability analysis of Vlasov equation.
The latter analysis shows that close to the unstable stationary
states of the Valsov equation, the relaxation occurs over a time
$\sim \ln N$ while close to stable stationary states, the system
stays in a quasistationary state whose lifetime goes as a power
law in $N$.

So far we considered only the deterministic dynamics. An
interesting direction would be to study such models when the
dynamical rules are stochastic. One such case has been discussed
in \cite{Mukamel:2005} where the system of Ising spins evolves via
microcanonical Monte Carlo dynamics. Since the Ising energy can be
obtained as a limiting case of the Hamiltonians considered here
(with anisotropy or on-site potential), it would be worthwhile to
study these systems with stochastic evolution rules. Besides, for
the HMF model, the ensemble equivalence in the equilibrium steady
state has been shown using large deviation method
\cite{Barre:2005}. Recent numerical studies
\cite{Morita:2004,Baldovin:2006a} of the dynamics of this system
in contact with a thermal bath have focused on the stable regime
$\epsilon > \epsilon^*$ and find that the quasistationary states
seen in the microcanonical ensemble survive but the lifetime
increases as a power law (in $N$) with an exponent that decreases
with increasing system-bath coupling. It would be interesting to
know if the unstable phase exhibits a similar dependence on the
coupling with the heat reservoir.

Acknowledgement: Financial support of the Israel Science
Foundation (ISF) is acknowledged. Visit of F.B. to the Weizmann
Institute has been supported by the Albert Einstein Minerva Center
for Theoretical Physics.


\end{document}

%% file: unstb2.tex
\begin{picture}(0,0)%
\includegraphics{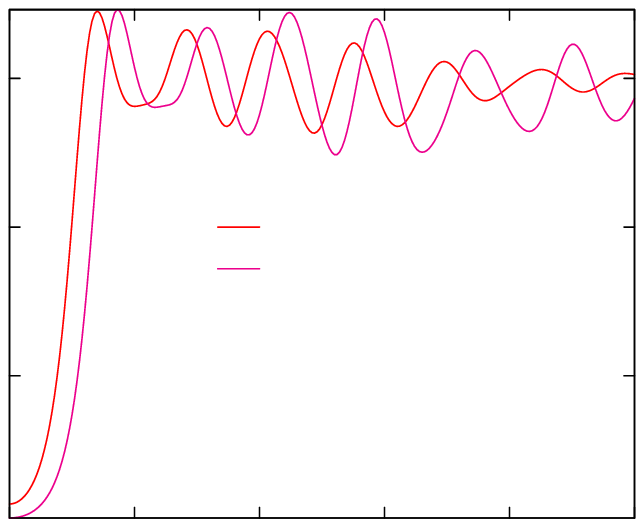}%
\end{picture}%
\begingroup
\setlength{\unitlength}{0.0200bp}%
\begin{picture}(12599,9719)(0,0)%
\put(2400,3848){\makebox(0,0)[r]{\strut{} 0.2}}%
\put(2400,5989){\makebox(0,0)[r]{\strut{} 0.4}}%
\put(2400,8130){\makebox(0,0)[r]{\strut{} 0.6}}%
\put(2700,1200){\makebox(0,0){\strut{} 0}}%
\put(4500,1200){\makebox(0,0){\strut{} 10}}%
\put(6300,1200){\makebox(0,0){\strut{} 20}}%
\put(8100,1200){\makebox(0,0){\strut{} 30}}%
\put(9900,1200){\makebox(0,0){\strut{} 40}}%
\put(11700,1200){\makebox(0,0){\strut{} 50}}%
\put(600,5460){\rotatebox{90}{\makebox(0,0){\strut{}$m(t)$}}}%
\put(7200,300){\makebox(0,0){\strut{}$t$}}%
\put(2880,8665){\makebox(0,0)[l]{\strut{}(a)}}%
\put(6600,5989){\makebox(0,0)[l]{\strut{}$N=10^3$}}%
\put(6600,5389){\makebox(0,0)[l]{\strut{}$N=10^4$}}%
\end{picture}%
\endgroup
 

%% file: unstb1.tex
\begin{picture}(0,0)%
\includegraphics{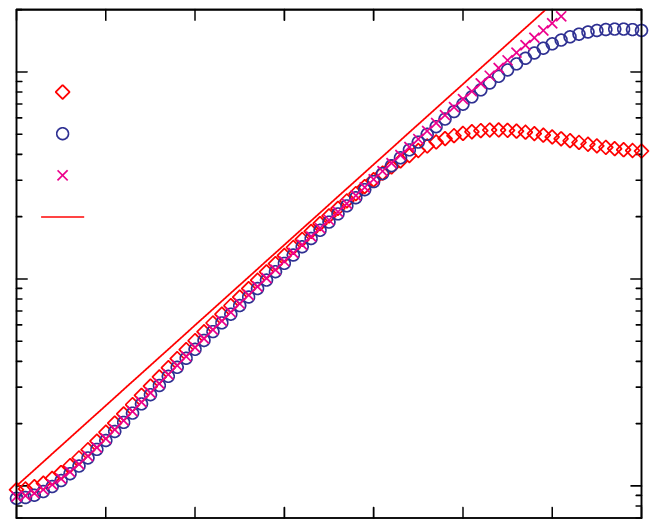}%
\end{picture}%
\begingroup
\setlength{\unitlength}{0.0200bp}%
\begin{picture}(12599,9719)(0,0)%
\put(2400,2262){\makebox(0,0)[r]{\strut{} 1}}%
\put(2400,5242){\makebox(0,0)[r]{\strut{} 10}}%
\put(2400,8223){\makebox(0,0)[r]{\strut{} 100}}%
\put(2700,1200){\makebox(0,0){\strut{} 0}}%
\put(3986,1200){\makebox(0,0){\strut{} 2}}%
\put(5271,1200){\makebox(0,0){\strut{} 4}}%
\put(6557,1200){\makebox(0,0){\strut{} 6}}%
\put(7843,1200){\makebox(0,0){\strut{} 8}}%
\put(9129,1200){\makebox(0,0){\strut{} 10}}%
\put(10414,1200){\makebox(0,0){\strut{} 12}}%
\put(11700,1200){\makebox(0,0){\strut{} 14}}%
\put(600,5460){\rotatebox{90}{\makebox(0,0){\strut{}$\sqrt{N} m_x(t)$}}}%
\put(7200,300){\makebox(0,0){\strut{}$t$}}%
\put(2861,8562){\makebox(0,0)[l]{\strut{}(b)}}%
\put(3964,7934){\makebox(0,0)[l]{\strut{}$N=10^4$}}%
\put(3964,7334){\makebox(0,0)[l]{\strut{}$N=10^5$}}%
\put(3964,6734){\makebox(0,0)[l]{\strut{}$N=5 \cdot 10^5$}}%
\put(3964,6134){\makebox(0,0)[l]{\strut{}${\rm exp}(t/\sqrt{5})$}}%
\end{picture}%
\endgroup
 

%% file: wU.tex
\begin{picture}(0,0)%
\includegraphics{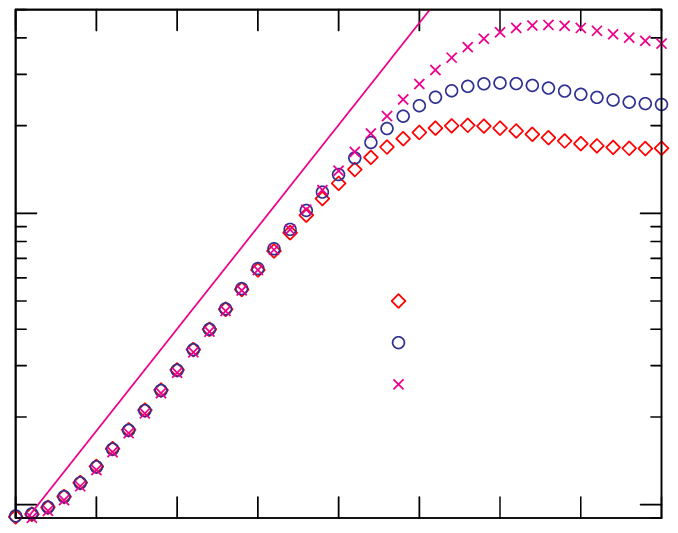}%
\end{picture}%
\begingroup
\setlength{\unitlength}{0.0200bp}%
\begin{picture}(12599,9719)(0,0)%
\put(2100,1992){\makebox(0,0)[r]{\strut{} 1}}%
\put(2100,6187){\makebox(0,0)[r]{\strut{} 10}}%
\put(2400,1200){\makebox(0,0){\strut{} 0}}%
\put(3563,1200){\makebox(0,0){\strut{} 1}}%
\put(4725,1200){\makebox(0,0){\strut{} 2}}%
\put(5888,1200){\makebox(0,0){\strut{} 3}}%
\put(7050,1200){\makebox(0,0){\strut{} 4}}%
\put(8213,1200){\makebox(0,0){\strut{} 5}}%
\put(9375,1200){\makebox(0,0){\strut{} 6}}%
\put(10538,1200){\makebox(0,0){\strut{} 7}}%
\put(11700,1200){\makebox(0,0){\strut{} 8}}%
\put(600,5460){\rotatebox{90}{\makebox(0,0){\strut{}$\sqrt{N} m_x(t)$}}}%
\put(7050,300){\makebox(0,0){\strut{}$t$}}%
\put(2691,8470){\makebox(0,0)[l]{\strut{}(a)}}%
\put(8513,4925){\makebox(0,0)[l]{\strut{}$N=10^3$}}%
\put(8513,4325){\makebox(0,0)[l]{\strut{}$N=2 \cdot 10^3$}}%
\put(8513,3725){\makebox(0,0)[l]{\strut{}$N=5 \cdot 10^3$}}%
\end{picture}%
\endgroup
 

%% file: wS.tex
\begin{picture}(0,0)%
\includegraphics{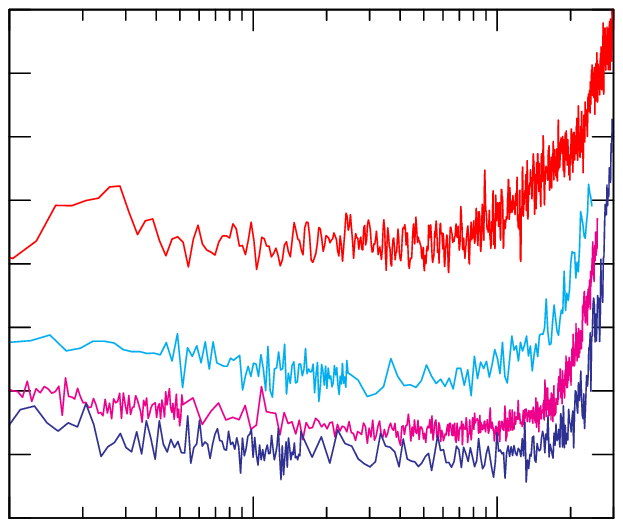}%
\end{picture}%
\begingroup
\setlength{\unitlength}{0.0200bp}%
\begin{picture}(12599,9719)(0,0)%
\put(2700,1800){\makebox(0,0)[r]{\strut{} 0}}%
\put(2700,2715){\makebox(0,0)[r]{\strut{} 0.02}}%
\put(2700,3630){\makebox(0,0)[r]{\strut{} 0.04}}%
\put(2700,4545){\makebox(0,0)[r]{\strut{} 0.06}}%
\put(2700,5460){\makebox(0,0)[r]{\strut{} 0.08}}%
\put(2700,6375){\makebox(0,0)[r]{\strut{} 0.1}}%
\put(2700,7290){\makebox(0,0)[r]{\strut{} 0.12}}%
\put(2700,8205){\makebox(0,0)[r]{\strut{} 0.14}}%
\put(2700,9120){\makebox(0,0)[r]{\strut{} 0.16}}%
\put(3000,1200){\makebox(0,0){\strut{} 1e-04}}%
\put(6512,1200){\makebox(0,0){\strut{} 0.001}}%
\put(10024,1200){\makebox(0,0){\strut{} 0.01}}%
\put(600,5460){\rotatebox{90}{\makebox(0,0){\strut{}$m(t)$}}}%
\put(7350,300){\makebox(0,0){\strut{}$t N^{-1.7}$}}%
\put(3340,8434){\makebox(0,0)[l]{\strut{}(b)}}%
\end{picture}%
\endgroup
 

%% file: gUp.tex
\begin{picture}(0,0)%
\includegraphics{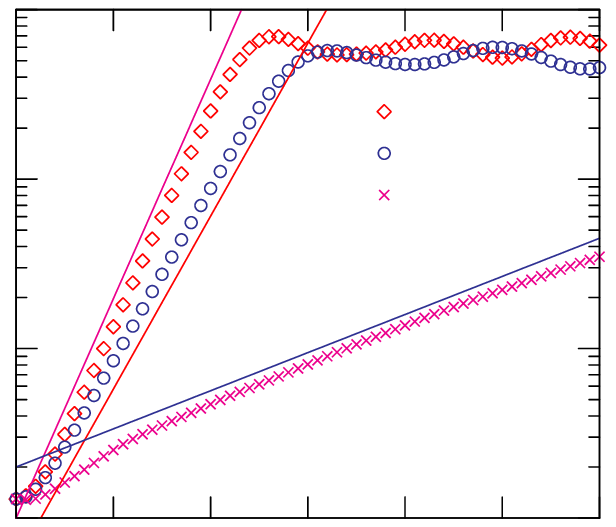}%
\end{picture}%
\begingroup
\setlength{\unitlength}{0.0200bp}%
\begin{picture}(12599,9719)(0,0)%
\put(3000,1800){\makebox(0,0)[r]{\strut{} 0.001}}%
\put(3000,4240){\makebox(0,0)[r]{\strut{} 0.01}}%
\put(3000,6680){\makebox(0,0)[r]{\strut{} 0.1}}%
\put(3000,9120){\makebox(0,0)[r]{\strut{} 1}}%
\put(3300,1200){\makebox(0,0){\strut{} 0}}%
\put(4700,1200){\makebox(0,0){\strut{} 2}}%
\put(6100,1200){\makebox(0,0){\strut{} 4}}%
\put(7500,1200){\makebox(0,0){\strut{} 6}}%
\put(8900,1200){\makebox(0,0){\strut{} 8}}%
\put(10300,1200){\makebox(0,0){\strut{} 10}}%
\put(11700,1200){\makebox(0,0){\strut{} 12}}%
\put(600,5460){\rotatebox{90}{\makebox(0,0){\strut{}$m_x(t)$}}}%
\put(7500,300){\makebox(0,0){\strut{}$t$}}%
\put(9200,7651){\makebox(0,0)[l]{\strut{}$\epsilon=0.55$}}%
\put(9200,7051){\makebox(0,0)[l]{\strut{}$\epsilon=0.75$}}%
\put(9200,6451){\makebox(0,0)[l]{\strut{}$\epsilon=1.5$}}%
\end{picture}%
\endgroup
 

%% file: paper.bbl
\begin{thebibliography}{10}
\bibitem{note} Relaxation times can be exponentially long if the initial
state is metastable  which we do not consider here.
\end{thebibliography}
